# Active individual nanoresonators optimized for lasing and spasing operation

A. Szenes[1], D. Vass[1], B. Bánhelyi[2], M. Csete[1,*]

1. *Department of Optics and Quantum Electronics, University of Szeged, Dóm tér 9, Szeged 6720, Hungary*
2. *Department of Computational Optimization, University of Szeged, Árpád tér 2, Szeged 6720, Hungary*

**Abstract**

Plasmonic nanoresonators consisting of a gold nanorod and a spherical silica-core and gold-shell, both coated by a gain layer, were optimized to maximize the near-field enhancement (NF-type) and far-field outcoupling (FF-type), and to enter into the spasing operation region (NF-c*-type). It was shown that in case of moderate concentration the nanorod has more advantages: smaller lasing threshold, larger slope efficiency and achieved intensities in the near-field, in addition in FF-type system smaller gain and outflow threshold, earlier flipping and larger far-field out-coupling efficiency. However, the near-field (far-field) bandwidth is smaller in for NF-type (FF-type) core-shell nanoresonators. In case of larger concentration although the slope efficiency and near-field intensity remain larger and the far-field redistribution is more efficient in case of the nanorod, the core-shell nanoresonator is more advantageous, taking into account the smaller lasing, outflow, absorption and extinction cross-section thresholds, as well as the larger internal and external quantum efficiencies. In addition to this the bandwidth of core-shell nanoresonator is also smaller. It was also shown that the strong-coupling of time-competing plasmonic modes accompanies the transition from lasing to spasing.

**Introduction**

Different types of metal nanoresonators are capable of supporting localized surface plasmon resonances (LSPR). Plasmonic nanorods have longitudinal and transversal resonances that can be excited via **E**-fields oscillating parallel to the longer and shorter axes, respectively. The lowest order dipolar resonances can reemit the energy into the far-field efficiently, whereas the higher order resonances are "dark" and dissipate energy into the near-field dominantly [1]. Dielectric-metal core-shell particles have a great potential to enhance the near-field due to the plasmon hybridization phenomenon [2]. Their resonance wavelength can be tuned through a wide band by varying their composition and generalized aspect ratio (GAR) [3]. The field enhancement inside such closed nanoresonators can be described analytically [4]. Due to improvement of the local density of states accompanying the near-field enhancement, plasmonic nanoresonators can prohibit or even inhibit the spontaneous emission of nearby fluorescent emitters [5]. The specific near-field confinement and far-field antenna effects depend on the proper tuning of the plasmonic mode, i.e. on the design of the coupled emitter-particle configuration [6, 7].

Plasmonic nanoresonators were also used to develop nanolasers that can deliver energy to the nanoscale on fs timescale and make it possible to achieve extremely high intensities in the electromagnetic near-field, which is important in high intensity laser physics as well as in intra-cavity spectroscopy [8, 9, 10, 11]. A seminal paper adopted the macroscopic criteria to nanolasers and uncovered that simultaneously desired is the low threshold of lasing and high threshold of bleaching. Analytical expressions were determined for the nanolasing threshold and slope efficiency, and the existence of an optimal concentration and Purcell factor for low threshold was shown. The increase of the concentration and Purcell factor is advantageous to delay bleaching, and larger confinement is advantageous for both thresholds [12].

To demonstrate the canonical lasing phenomena in plasmonic systems novel numerical approaches were developed. The steady-state simulation enabled to distinguish three regions, where the gain material integrated into a plasmonic crystal behaves as an absorber, optical amplifier or a laser. To extract the optical response of the system, the coupling between the **E**-field and the population (induced polarization) was formulated. The effect of the field intensity onto the population inversion was taken into account through a coupling term, and the gain medium was described by a complex permittivity including the local absorption and gain coefficients [13]. The spatio-temporal modelling of lasing in time-domain generalization of FEM was based on the solution of the field equation formulated for the vector potential, which includes the time-dependent polarization corresponding to stimulated absorption and emission. In a gain medium consisting of four-level emitters the dynamics was described by a Lorentzian equation including the coupling term between the polarization and the **E**-field, and the populations were described by the rate equations. It was shown that the lasing threshold depends on the parameters of the gain medium and the plasmon resonance enabling the enhancement, which can be optimized [14]. Maxwell-Bloch Langevin time-domain equations were solved, which include information about the resonance frequencies and coupling constants specific for the gain material. In addition, a noise term played as a seed aiding the transient build-up of the coherent lasing fields through feedback that forces stimulated emission in the gain material. It was shown that above threshold oscillations develop in-time, and line-width narrowing occurs in frequency, whereas below threshold only amplified spontaneous emission (ASE) is achievable [15].

Time-domain simulation of the nanolaser dynamics can be a resource and time consuming process because of the ps-ns timescales of the population transitions. To simulate the spontaneous emission an artificial source was introduced by placing random phase dipole sources at FDTD mesh points, whereas the rate equation modelling of the stimulated emission was realized by considering the slowly varying saturated state as an initial state. This way it was proven that both emission pathways can be enhanced, when the emission frequency is coincident with the plasmonic resonance [16].

Loss compensation via metamaterials was also investigated by determining the steady-state occupations via CW excitation at the absorption, and by solving the time-dependent rate equations with numerical pump-probe simulations. By taking into account that the spatially inhomogeneous gain is the function of the local pump **E**-field it was shown that loss compensation is achieved by good overlap between the emission and plasmonic band. To overcome radiative and dissipative losses and to reach lasing a higher dye concentration was proposed [17]. In case of coexistent bright and dark modes the lasing of a bright mode was demonstrated, when the higher-Q dark mode was discriminated spectrally and spatially. It was shown that by changing the spectral alignment of the emission line or the spatial deposition of the gain, the relative intensities of the modes and the limits, where either switches off, can be manipulated [18, 19]. Even more accurate realistic description was achieved with frequency-domain FDTD simulations by taking into account the attenuation of the pump intensity, namely by incorporating the local field dependent absorptance at the pump wavelength and the inhomogeneous gain distribution at the probe wavelength as well [20]. The possibility of lasing improvement by plasmonically enhancing the pump phenomenon was also demonstrated [21].

A novel spasing phenomenon was discovered, where the plasmonic modes are involved into stimulated emission. The signatures of laser action such as the appearance of a threshold and linewidth narrowing was experimentally demonstrated in the simple plasmonic lasing system based on a gold nanosphere coated by a dielectric layer containing dye molecules [22]. It was shown that the high threshold in case of the first described gain-shell coated metal nanosphere type nanolasers is caused by the interband transition of gold nearby the emission, which can be decreased by optimizing the lasing wavelength, the nanosystem geometry and by increasing the background index [23].

Plasmonic loss compensated by optical gain has been reported in case of a nanorod coated with dye doped dielectric layer. It was shown that the emission wavelength can be tuned by changing the doping level and chemical composition. By increasing the dye concentration the emission wavelength redshifts, threshold becomes larger, but the intensity enhancement becomes smaller. In presence of nanorod more significant linewidth narrowing was observed with particles than with passive dielectric coating only, in addition to this ultrasmall mode volume ad ultrahigh Purcell factor was achieved simultaneously [24].

The weak-form solution of the field equation formulated for the vector potential enabled to perform the dynamical examination of the nanolasing phenomenon in FEM. The seeding of the emission channel was initialized by a small probe signal. Detailed inspection of the threshold behavior (that is inversely proportional to the spatial confinement factor) and slope efficiency (that is proportional to the Q factor) revealed that an optimal geometry of nanorod exists, which allows to maximize confinement both in space and time [25].

It was shown that the spasing occurs at the threshold value of the frequency and gain, whereas the achieved gain depends on the local field strength accompanying the pump. In case of synchronization in core-shell nanoparticles a homogeneous internal field appears for vanishing external field as well. According to the existence of poles in the nanoparticles polarizability both the field strength and the scattering cross-section may diverge (the latter is accompanied by negative absorption and zero extinction). As a consequence, spasers' linear modelling may result in unphysical phenomena, that can mitigated by including the saturation into the complex dielectric function [26]. For both of metal-gain and gain-metal core-shell particles it was stated that a minimum gain threshold exists, the wavelength of which depends only on the dispersion of the bounding materials. The geometry tuning makes it possible to achieve resonance via different N shape-mode factors. However, the statement regarding the equal effectiveness of any nanoparticle shape that has the value of N resulting in resonance at the wavelength of minimal gain" might need revision, by considering the (i) slope efficiency, (ii) near-field confinement factor, (iii) far-field extraction efficiency simultaneously. It was shown that in case of nanorods large aspect ratios are appropriate to achieve the wavelength corresponding to minimal threshold, whereas in case of core-shells the resonance appears at a wavelength determined by the average of the bounding media [27]. It was also demonstrated that in case of a hollow spherical plasmonic nanoparticle consisting of an oblate gain-assisted core by increasing the aspect ratio the threshold can be decreased, the achieved absorptance and scattering can be increased by several orders of magnitude, and when the extinction becomes zero, the linewidth strongly decreases. This is due to the high quality of superresonance for oblate cores that promotes the spasing [28]. Different geometries were compared to consider the possible advantages for near- and far-field operation. It was shown that the light extraction efficiency is larger and decays slowly in oblate spheroids that are ideal for lasing, whereas it is lower and rapidly decreasing in prolate spheroids, that are more appropriate for local electromagnetic field enhancement [29].

There is a consensus regarding that a laser threshold criterion has to be reconsidered in case of nanolasers. The "S" curve, which is the intensity as a function of pump on log-log scale, does not work, because of ideal nanolasers have β=1 (~unity stimulated rate / spontaneous rate), i.e. the kink disappears, so they can be considered as threshold-less lasers [30]. High-intensity, coherence time, special photon autocorrelation functions are important criteria to conclude about systems, whether they emit a laser light. This can be evidenced by examining the linewidth narrowing, the second-order autocorrelation function, which has to exhibit $g^2(0)$: 2->1 modification at the transition from thermal to coherent emission [31].

In the latest reviews the main design rules of spasers were summarized as follows: the threshold of a spaser is governed by the dielectric properties and the quality factor, but does not depend directly on the geometry.

The threshold-less lasing corresponds to β=1, the minimal threshold corresponds to the case, when the mode rate is much larger than the loss in gain, and the pump approximates the mode decay rate. The typical threshold is 1-100 MW/m$^2$, whereas the external quantum efficiency (EQE) is ~10% [32]. Nanolasers and spasers have promising properties for the applications as multifunctional optical biological probes [10, 11, 33, 34]. They do not saturate and they are more resistant to photobleaching. Molecule specific low-toxicity ultrafast probes with a narrow spectrum and bright emission was reported. The spasers have a great potential for photothermal cancer therapy and photoacoustic imaging [33, 34]. Ultranarrow linewidth and population depletion of spasers can be utilized in superresolution microscopy. Recently ultranarrow spatial resolution is achieved with a few nm linewidth of a three level dye doped coated plasmonic nanoparticle in stimulated emission depletion (STED) imaging [35, 36].

**Methods**

Based on the literature frequency-domain approach is justified by the fact that in many cases the properties of the equilibrium, i.e. of the steady-state operation, are more interesting than the transient phenomena. The advantage of this approach is that the amplifying medium is simply described by a phenomenological permittivity or refractive index with a negative imaginary part at the emission. Moreover, optimization of the geometry and configuration for any time-dependent objective function is a difficult task for nanolasers and spasers, hence optimization in the frequency-domain may be more expedient. The post-evaluation of the dynamics in the optimized nanolasing system makes it possible to consider further potential advantages as the the dominance of the transient modes that are at play.

Plasmonic nanoresonators of metal-gain and dielectric-metal-gain composition were optimized to construct plasmonic nanolasers. The former system is a gold nanorod (NR) embedded into a gain medium, whereas the latter is a spherical silica-gold core-shell (CS) nanoparticle coated by a gain layer. The gain medium was a polymer embedding rhodamine (Rh800) dye molecules that were considered as four-level emitters with an absorbing and an emitting Lorentz oscillator at 680 nm and 710 nm, respectively.

Numerical pump and probe experiments were performed in frequency-domain finite element (FEM) computations realized by the RF module of COMSOL Multiphysics. A three step study was realized: (i) pump beam was launched at the absorption wavelength of the dye to initiate the population inversion, (ii) a weak probe pulses at the emission wavelength with an order of magnitude smaller intensity was used to initiate the primary photons for stimulated emission enhancement, (iii) the passive system without the plasmonic nanoresonators was also inspected. The number of dye molecules on different levels as well as the near-field intensity was monitored. The gain material was described via the usual rate equations including two stimulated transitions, as well as the radiative and non-radiative spontaneous transitions [13, 15, 20].

These rate equations include the polarization density ($P_i$) and the local electric field ($E_i$) strength at the absorption ($\omega_a$) and emission ($\omega_e$) wavelength as well,

$$\frac{\partial N_3}{\partial t} = \frac{1}{\hbar \omega_a}\left(\frac{\partial \boldsymbol{P_a}}{\partial t} + \frac{\Delta\omega_a}{2}\boldsymbol{P_a}\right)\boldsymbol{E_a} - \frac{N_3}{\tau_{32}}, \qquad (1)$$

$$\frac{\partial N_2}{\partial t} = \frac{1}{\hbar \omega_e}\left(\frac{\partial \boldsymbol{P_e}}{\partial t} + \frac{\Delta\omega_e}{2}\boldsymbol{P_e}\right)\boldsymbol{E_e} - \frac{N_2}{\tau_{21}} + \frac{N_3}{\tau_{32}}, \qquad (2)$$

$$\frac{\partial N_1}{\partial t} = -\frac{1}{\hbar \omega_e}\left(\frac{\partial \boldsymbol{P_e}}{\partial t} + \frac{\Delta\omega_e}{2}\boldsymbol{P_e}\right)\boldsymbol{E_e} - \frac{N_1}{\tau_{10}} + \frac{N_2}{\tau_{21}}, \qquad (3)$$

$$\frac{\partial N_0}{\partial t} = -\frac{1}{\hbar \omega_a}\left(\frac{\partial \boldsymbol{P_a}}{\partial t} + \frac{\Delta\omega_a}{2}\boldsymbol{P_a}\right)\boldsymbol{E_a} + \frac{N_1}{\tau_{10}}, \qquad (4)$$

and the time-dependent $\boldsymbol{P}_i$ polarization vector field coupling to the $\boldsymbol{E}_i$ -field was specified as follows:

$$\frac{\partial^2 \boldsymbol{P_i}}{\partial t^2} + \Delta\omega_i \frac{\partial \boldsymbol{P_i}}{\partial t} + \omega_i^2 \boldsymbol{P_i} = -\sigma_i \Delta N_i \boldsymbol{E_i}, \qquad (5)$$

where i=a,e refers to the absorptance and emission frequency respectively, $\tau_{10}$, $\tau_{21}$ and $\tau_{32}$ are the lifetimes of the transitions, $\Delta\omega_a$ and $\Delta\omega_e$ are the FWHM of the corresponding spectral lines, $\sigma_a$ and $\sigma_e$ is the coupling constant between the polarization and electric field, $\Delta N_a = N_3 - N_0$ and $\Delta N_e = N_2 - N_1$ are the population differences, whose steady state values can be determined as follows [20]:

$$\Delta N_a = \frac{\tau_{32}\Gamma_{03,eff} - 1}{1 + (\tau_{32} + \tau_{21} + \tau_{10})\Gamma_{03,eff}} N, \qquad (6)$$

$$\Delta N_e = \frac{(\tau_{21} - \tau_{10})\Gamma_{03,eff}}{1 + (\tau_{32} + \tau_{21} + \tau_{10})\Gamma_{03,eff}} N, \qquad (7)$$

where N is the total number of dye molecules and $\Gamma_{03,eff} = \Gamma_{03,eff}(\boldsymbol{E}_{local})$ is the time-independent (steady-state) effective probability of the 0-to-3 stimulated transition. Based on the spatially inhomogeneous population difference, the local intensity dependent complex dielectric permittivity of the gain material at the pump and probe frequency can be defined in the following general form [20]:

$$\varepsilon(\omega) = \varepsilon_{host} + \sum_i \frac{1}{\varepsilon_0} \frac{\sigma_i \Delta N_i}{\omega^2 + i\omega\Delta\omega_i - \omega_i^2}, \qquad (8)$$

assuming that both the absorption and emission bands have a Lorentz lineshape. To account for the saturation and to avoid numerical artefacts, a local **E**-field dependent nonlinear term was included into the denominator of Eq(8) as well [27, 29].

The population inversion, the complex dielectric permittivity and refractive index were determined as a function of normalized external field of $\boldsymbol{E}_{pump}/\boldsymbol{E}_{sat}$ (referred as $E_{sat}^{pump}$) and normalized internal field of $\boldsymbol{E}_{local}/\boldsymbol{E}_{sat}$ (referred as $E_{sat}^{local}$), where field strength corresponding to saturation is $E_{sat} = \sqrt{4\hbar\omega_a \Delta\omega_a/(\tau_{21}\sigma_a)}$, which takes on a value of 2.94*10$^6$ V/m for the Rh800 dye. The existence of double **E**-field scales is relevant according to the considerable enhancement of the local **E**-field by the plasmonic nanoresonators. The **E**-field enhancement at the pump wavelength was determined based on the relation of these two scales $\boldsymbol{E}_{local}/\boldsymbol{E}_{pump}$, (referred as $E_{sat}^{local}/E_{sat}^{pump}$) (*See supplementary material, Table S1*).

The optimization of the nanoresonator geometries was realized by using two different objective functions: (a) maximal near-field enhancement, by monitoring the maximal as well as the average **E**-field inside the gain medium at the lasing frequency (NF-c nanoresonators) (Fig. 1); (b) maximal far-field enhancement at the lasing frequency, by monitoring the absorption inside the metal and the gain medium, as well as the power outflow and its polar angle distribution (FF-c nanoresonators) (Fig. 2).

The lasing-threshold ($\boldsymbol{E}_{th}$) and the slope efficiency was determined based on the maximal and average **E**-field as a function of the local and pump intensity. The Ohmic loss in the metal and gain medium, and the power outflow were also determined, and the internal (IQE) and external (EQE) quantum efficiency was evaluated. In order to prove that the stimulated emission is enhanced via the coupled nanoresonators the FWHM of the spectral distribution of the near-field enhancement and far-field emission was inspected, and the line-width narrowing was concluded based on comparison with the full-width-at-half-maximum (FWHM) of corresponding spectral distributions in the passive systems.

Based on the previous literature at a specific dye concentration there always should exist two sets of geometry parameters, which promote maximal near-field enhancement and efficient out-coupling, respectively. However, the near-field maximization does not ensure inherently that the coupled system enters into a spasing region, where the large gain is converted into enhanced scattering and the extinction can be completely eliminated [26-29]. Therefore, for the systems optimized for near-field maximization the effect of the gain concentration increase on the steady-state and time-dependent response was inspected as well (NF-c* nanoresonators) (Fig. 3).

The underlying near-field phenomena were uncovered by studying the accompanying charge and near-field distribution and by inspecting the polar angle distribution of the far-field emission as well (Fig. 4). The NF and FF, as well as the nanorod and core-shell nanoresonators were compared to consider the advantages of the different systems.

**Results**

Compared to the maximum the average **E**-field is significantly smaller and less rapidly increases by increasing the pump intensity in all inspected systems, proving the inhomogeneity of the field distribution (Fig. 1, 2/a). The absorptance in the metal and gain medium have a similar "S" shape as the local **E**-fields (Fig. 1, 2/b). In all active nanoresonators the negative absorptance in the gain is distributed partially in the competitive loss channel of gold absorptance and in the outflow, the latter is comparable to the total power of the probe beam. Evaluation of the optical cross-sections (OCS) makes it possible to prove the advantages of the NF-c and FF-c active nanoresonators in the achievement of large near-field enhancement and large out-coupling efficiency, respectively (Fig. 1-2/c). The maximum in the near-field spectrum appears around the emission wavelength, which proves that the optimized systems are capable of enhancing the emission (Fig. 1-2/d). The linewidth narrowing proves enhanced stimulated emission, i.e. lasing-like behaviour in all of the NF and FF optimized nanoresonators (Fig. 1,2/d, e). The pump intensity dependent tendencies unambiguously prove that the NF-c nanoresonators confine the **E**-field sufficiently, whereas the FF-c nanoresonators couple to the far-field efficiently and also redistribute the beam spatially (Fig. 1, 2/d-f).

*Nanorod and core-shell mediated near-field amplification: NF-c systems*

According to the objective function that was defined to achieve the largest average near-field enhancement, the optimized nanorod (NF-NR-c) of 59.38 nm*23.39 nm long and short axis (AR= 2.54) with a gain-coating of 18.13 nm thickness is significantly smaller in all dimensions than its FF-NR-c counterpart, which shows that it is better suited to absorb the incoming probe light rather than to scatter it (Fig. 1a-to-2a insets and Table S1.). The ratio of the internal local and external pump fields ($\boldsymbol{E}_{sat}^{local}/\boldsymbol{E}_{sat}^{pump}$) at the population saturation reveals ~54-fold near-field enhancement (Fig. S1a, Table S1.). The dielectric-metal-gain multilayer composition optimization of the same objective function resulted in a core of 18.99 nm radius, a metal shell of 5 nm and a gain medium of 24.62 nm thickness, respectively (NF-CS-c), all are significantly smaller than the corresponding parameters in the FF-CS-c counterpart (Fig 1a-to-2a insets and Table S1). The ratio of the local and pump fields ($\boldsymbol{E}_{sat}^{local}/\boldsymbol{E}_{sat}^{pump}$) at the population saturation reveals ~32-fold near-field enhancement (Fig. S1a, Table S1). The near-field enhancement at the pump wavelength extracted from the ratio of saturation thresholds is smaller for NF-CS-c than for NF-NR-c.

The achieved gain affects noticeably $n(\omega_a)$ and significantly $\varepsilon_{im}(\omega_a)$ and $\kappa(\omega_a)$ with respect to the host medium at the pump wavelength (Fig. S1/b, d). Already at small pump positive imaginary values are taken on, accordingly there is a nonzero absorptance at the pump wavelength, which decreases and converges to zero by increasing the pump. The intervals, where the $n(\omega_a)$, $\varepsilon_{im}(\omega_a)$ and $\kappa(\omega_a)$ values are taken on are slightly larger for NF-CS-c, than for NF-NR-c (Fig. S1/b, d and Table S1).

There is a considerable deviation both in the real and imaginary parts from the passive system permittivity and refractive index at the probe frequency (Fig. S1/c, e). In case of weak pump, the small positive $\varepsilon_{im}(\omega_e)$ and $\kappa(\omega_e)$ values indicate that the gain medium is still weakly absorbing, but above a certain pump the switching in sign proves that a gain is settled on, then the imaginary parts are gradually decreasing. The intervals, where the imaginary part of permittivity and refractive index values are taken on, are slightly larger for NF-CS-c (Fig. S1/c, e and Table S1).

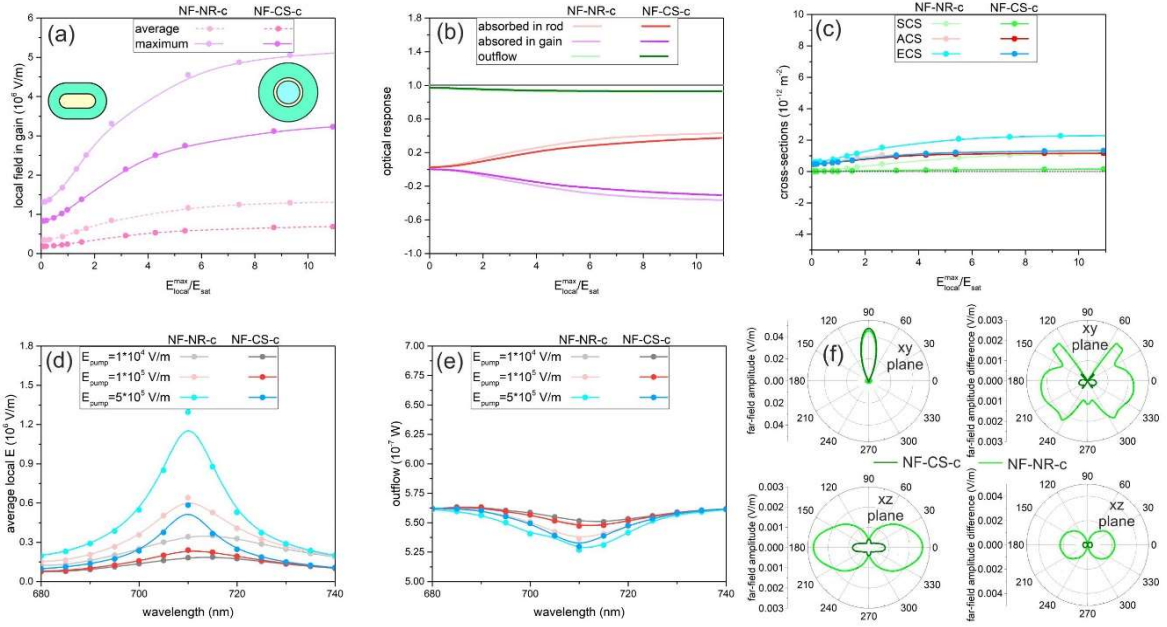

Figure 1. Normalized pump **E**-field dependent optical response of NF-NR-c and NF-CS-c optimized nanoresonators: (a) average and maximum of the enhanced local probe **E**-field in the gain medium, (b) absorptions in the metal nanorod and shell and in the gain medium, and the far-field emission, (c) optical cross-sections as a function of pump amplitude, spectra of the (d) average near-field and (e) far-field power outflow, (f) polar angle distribution in the far-field. Inset: geometry of NF-NR-c and NF-CS-c nanoresonators.

In NF-NR-c after a transitional small slope region the local probe **E**-fields are linearly increasing throughout [$E_{sat}^{local}$: 0.31-2.63] and then saturate [2.63<$E_{sat}^{local}$] (Fig.1a, Table S1). The extrapolated lasing threshold in the internal local (external pump) **E**-field normalized to the $E_{sat}$ is $E_{sat}^{local}$: 0.30 ($E_{sat}^{pump}$: 0.007). The (2.16*10$^5$ V/m) slope in the average probe **E**-field is considerably smaller than in the probe **E**-field maximum (8.52*10$^5$ V/m). In NF-CS-c after the extrapolated threshold in the internal local (external pump) **E**-field normalized to $E_{sat}$ of $E_{sat}^{local}$: 0.43 ($E_{sat}^{pump}$: 0.013), the near-field increases linearly in the interval of [$E_{sat}^{local}$: 0.48-3.16] with a slope of 9.8*10$^4$ V/m in the average probe **E**-field and with a slope of 4.97*10$^5$ V/m in the maximum of probe **E**-field. In NF-CS-c the linear region is wider, the lasing threshold is larger, whereas all of the slope efficiency and the achieved average and maximum probe **E**-field values are smaller than in NF-NR-c.

In NF-NR-c the gain absorptance is positive at the transitional region of small pump rates, then becomes negative at the gain-threshold values of $E_{sat}^{local}$: 0.32 ($E_{sat}^{pum}$: 0.006). In contrast, in NF-CS-c the threshold of negative gain absorptance is $E_{sat}^{local}$: 0.31 ($E_{sat}^{pump}$: 0.01). The threshold of negative gain is slightly smaller (larger) in the internal local (external pump) **E**-field in NF-CS-c similarly to the saturation threshold.

The positive metal absorptance overrides the negative gain absorptance throughout the complete inspected pump interval, as a consequence there is no far-field enhancement, neither in NF-NR-c nor in NF-CS-c (Fig. 1b). The absorptances in NF-CS-c are slightly smaller, whereas the outflow values are almost the same in the two optimized NF-c systems.

Accordingly, in both NF-c nanoresonators the absorption cross-section (ACS) is gradually increasing and remains positive, which indicates an uncompensated loss (Fig. 1c). In NF-NR-c the gradually increasing scattering cross-section (SCS) becomes commensurate with the absorption cross-section, as a results the extinction cross-section (ECS) more rapidly increases.

In contrast, in NF-CS-c the scattering cross-section remains very small throughout the inspected pump intensity interval, hence the slowly and monotonously increasing absorption and extinction cross-sections are close to each other. As a result, all optical cross-sections are larger in NF-NR-c than in NF-CS-c. By increasing the pump intensity, the peak on the spectrum of the average **E**-field around the passive resonance is slightly blue-shifted, becomes more intense, and its FWHM decreases (Fig. 1d). The near-field spectra of NF-CS-c and NF-NR-c are similar but more than two-times less intense in NF-CS-c. In case of NF-NR-c in the near-field spectrum there is no linewidth narrowing at small pump intensities, then a gradual 3.4-fold decrease begins and a 13.8 nm linewidth is reached. Similar tendency is observable in NF-CS-c, as a result a slightly larger (3.6-fold) decrease is achieved with a larger rate from smaller FWHM of the passive nanoresonator to a smaller 12.4 nm linewidth of the active nanoresonator (Fig. 1d, Table S1).

At the same time the dip of the outflow spectrum around the passive resonance deepens and undergoes a linewidth narrowing by increasing the pump intensity for both NF nanoresonators (Fig. 1e). In case of the NF-NR-c beside the primary line-width narrowing at larger pump a secondary dip appears at a smaller wavelength, which reveals the existence of coupled modes (Fig. 1e, turquoise). Although, the peaks on the outflow spectra of the passive systems are much narrower than the FWHM of the near-field peaks, they decrease with a smaller rate (~1.5) to considerably larger remaining linewidths (~19 nm) both in NF-NR and NF-CS-c active systems. This is in accordance with that NF systems are optimized to improve near-field properties (Fig. 1f, Table S1).

In contrast, in the spectrum of NF-CS-c there is still no sign of mode competition in the inspected pump interval (Fig.1e, blue). For further details, see the section about NF-NR-c* nanoresonators.

The polar diagram of far-field radiation enhancement shows that there is no radiation enhancement in the direction of probe propagation (90°). In the complementary polar angle regions, the NF-NR-c amplifies the radiation by redistributing it in space nearly uniformly. In contrast NF-CS-c enhances the radiation in four distinct directions with a considerably smaller intensity (Fig. 1f).

*Nanorod and core-shell mediated lasing: FF-systems*

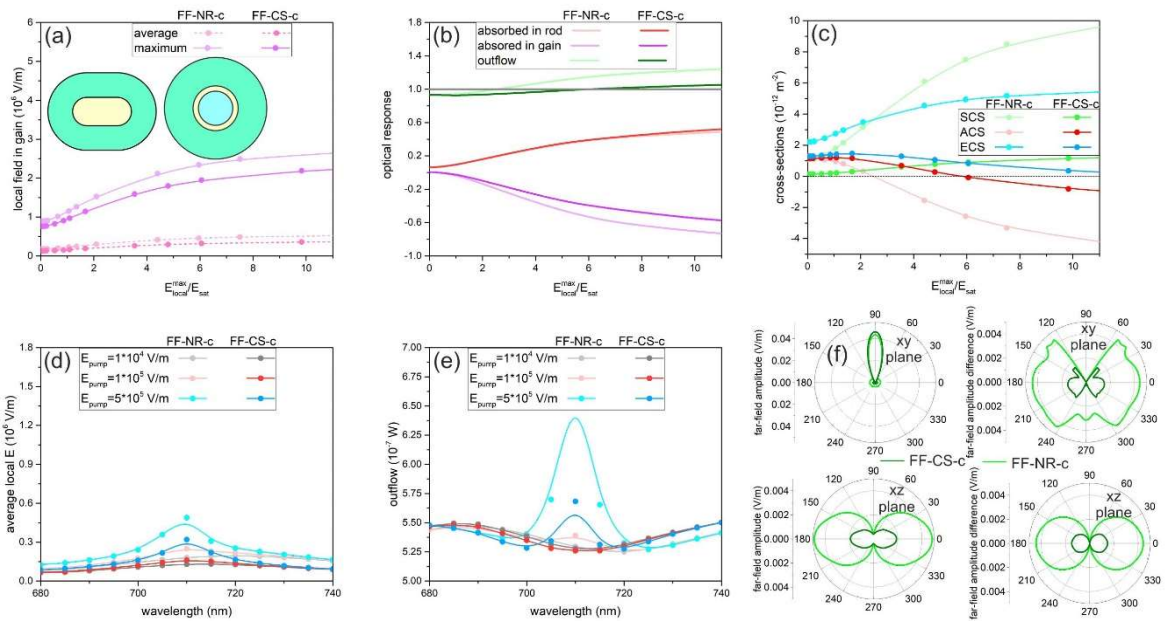

Figure 2. Normalized pump **E**-field dependent optical responses of FF-NR-c and FF-CS-c optimized nanoresonators: (a) average and maximum of the enhanced local probe **E**-field in the gain medium, (b) absorptions in the metal nanorod and shell and in the gain medium, and the far-field emission, (c) optical cross-sections, spectra of the (d) average near-field and (e) far-field power outflow, (f) polar angle distribution in the far-field. Inset: geometry of the FF-NR-c and FF-CS-c nanoresonators.

In case of the nanorod the optimization realized to enhance the far-field stimulated emission (FF-NR-c) the 97.33 nm*47.39 nm nanorod having a smaller aspect ratio of 2.05 was determined, that is coated by a much thicker 41.33 nm gain layer. This indicates that the FF-NR-c is more sphere-like, and the significantly larger size of both axes results in that both the gold nanorod and the gain volume is an order of magnitude larger, compared to the NF-NR-c (Fig. 2a-to-1a insets and Table S1). The dielectric-metal-gain multilayer composition optimization realized with the same objective function resulted in a larger core of 28.47 nm radius, a metal shell and a gain medium of 8.85 nm and 47.45 nm thickness, respectively (FF-CS-c) (Fig 2a-to-1a insets and Table S1). These parameters correspond to a five-fold larger gain and two-fold larger metal volume. The larger size proves that both the FF-NR-c and the FF-CS-c are capable of promoting far-field scattering, rather than near-field enhancement and metal absorption.

In FF-NR-c the population inversion exhibits a slower saturation at $E_{sat}^{local}$: 3.89 ($E_{sat}^{pump}$: 0.09), the ratio of the internal local (external pump) **E**-field strengths reveals a 43.22-fold near-field enhancement, which is smaller than in case of NF-NR-c (Fig. S1a, Table S1). The $E_{sat}^{local}$: 4.11 ($E_{sat}^{pump}$: 0.11) internal local (external pump) field-strengths corresponding to saturation indicate 37.36-fold near-field enhancement in FF-CS-c system that is larger than in NF-CS-c (Fig. S1a, Table S1). Both the internal local and external pump field-strength values corresponding to saturation are larger, but the near-field enhancement at the pump wavelength extracted from the ratio of saturation thresholds is smaller in case of FF-CS-c than in FF-NR-c. This indicates that the pump process is less enhanced in the FF-CS-c nanoresonator than via FF-NR-c, similarly to counterpart NF nanoresonators.

Similar permittivity and index of refraction is achieved in FF-NR-c and FF-CS-c, after a (faster-to-slower and faster) slightly faster saturation (of imaginary and real parts in FF-NR-c) in FF-CS-c compared to their NF counterparts, respectively (Fig. S2/b, e). Only the $\varepsilon_{imag}(\omega_a)$ and $\varepsilon_{imag}(\omega_e)$ values modify in a slightly smaller and larger interval in the FF-NR-c than in the FF-CS-c (Fig. S2/b, c, Table S1).

However, there are more significant differences between the near-field and far-field efficiencies of the FF optimized nanoresonators according to their different objective functions. In FF-NR-c, the $E_{sat}^{local}$: 0.32 ($E_{sat}^{pump}$:0.007) lasing thresholds are just slightly larger, the 6.83*10$^4$ V/m slope efficiency in the linear region of the average probe **E**-field is ~3-times smaller than in NF-NR-c indicating a weaker near-field confinement. The achieved maximum and average probe **E**-field is approximately and more than two-times lower, respectively (Fig.2a-to-1a and Table S1). In FF-CS-c the $E_{sat}^{local}$: 0.34 ($E_{sat}^{pump}$:0.009) lasing thresholds are considerably smaller, the 4.5*10$^4$ V/m slope efficiency in the linear region and the achieved maximum and average probe **E**-field is more than and almost two-times smaller than in NF-CS-c (Fig.2a-to-1a and Table S1). The FF-NR-c allows smaller lasing threshold, larger slope efficiency and larger near-field enhancement at the probe wavelength than FF-CS-c. However, the relation between FF-NR-c and FF-CS-c is similar to their NF counterparts in all these quantities.

Despite the significantly smaller NF enhancement and the larger volume of FF-NR-c, the NF and FF optimized nanorod absorptance values are surprisingly close to each other at the emission (Fig.2b-to-1b). In FF-NR-c the gain absorptance becomes negative at the gain-threshold values of $E_{sat}^{local}$: 0.32 ($E_{sat}^{pum}$: 0.007), whereas in NF-CS-c the threshold of negative gain absorptance is $E_{sat}^{local}$: 0.33 ($E_{sat}^{pump}$: 0.009). In FF-c nanoresonators the gain-threshold values are very similar to those in NF-c nanoresonators, however the thresholds of negative gain become slightly larger in NF-CS-c, similarly to the saturation thresholds.

Even if the Ohmic loss is similar in the NF and FF nanorods, in FF-NR-c the larger gain makes larger power outflow possible than in the passive system above the outflow-threshold of $E_{sat}^{local}$=2.42.

As a result, significant pump energy is converted into far-field radiation. In contrast, the absorptance value in FF-CS-c is more enhanced with respect to its NF-CS-c counterpart, but the larger gain allows a power outflow enhancement at the $E_{sat}^{local}$=6.1 outflow-threshold (Fig. 2b-to-1b, Table S1). The more rapidly increasing and significantly larger gain in FF-NR-c makes it possible to better enhance the far-field radiation, moreover the outflow-threshold shows up at significantly smaller pump than in FF-CS-c.

In FF-NR-c the effect of the overcompensated absorption is observable in the pump dependent cross-section tendency as well, since the absorption cross-section changes from positive to negative values at the ACS-threshold of $E_{sat}^{local}$: 2.48, with increasing absolute values as the pump increases (Fig. 2c, Table S1). However, caused by the large scattering cross-section, the extinction remains positive and even monotonously increases throughout the inspected pump range, very similarly to the NF-NR-c system. The larger cross-section values show that much more dynamic energy transitions occur in FF-NR-c than in NF-NR-c, again demonstrating a significant difference between the NF and FF optimal systems (Fig. 2c-to-1c).

The initial cross-sections are significantly larger in FF-CS-c than in NF-CS-c. However, while the scattering cross-section monotonously increases and remains higher throughout the inspected pump interval, the absorption and extinction cross-sections exhibit a global maximum, and then rapidly decrease. As a result, the absorption cross-section becomes negative, however at a considerably larger local pump field strength (ACS-threshold of $E_{sat}^{local}$: 5.77), than in FF-NR-c (Fig. 2c, Table S1). In FF-CS-c the negative absorption cross-section almost compensates the positive contribution of the scattering cross-section, and the extinction cross-section approaches zero. As a result, all optical cross-sections are larger in NF-NR-c than in NF-CS-c, except the absorption cross-section. These results indicate that while the FF-NR-c system is far from the spaser operation region, the FF-CS-c approximates it close to the upper bound of the inspected pump interval (Fig. 2c-to-1c).

By increasing the pump, gradually a more intense near-field is reached, and on the spectrum a linewidth narrowing is observable on a slightly blue-shifting peak both in FF-NR-c and FF-CS-c (Fig. 2d). The amplitude of the near-field spectral peak around the passive resonance is significantly larger in FF-NR-c than in FF-CS-c. The near-field spectra of FF-NR-c and FF-CS-c show linewidth broadening at small pumps, then the FWHMs start to gradually decrease with increasing pump intensity (Fig. 2d). The 6.1-fold decrease is much larger than in NF-NR-c and results in a smaller FWHM (12 nm) despite the wider passive spectrum (Table S1). In FF-CS-c the 4.1-fold decrease is still considerably larger than in NF-CS-c however the linewidth remains slightly larger (12.9 nm) at the largest inspected pump intensity. Accordingly, the linewidth of near-field spectrum of FF-NR-c is always larger than in FF-CS-c except at the largest pump where they are comparable.

The far-field spectrum of FF-NR-c and FF-CS-c exhibits a peak instead of a dip at two consecutive and at the highest applied pump intensity, respectively (Fig. 2e). While both the near-field and outflow spectra are similar in FF-NR-c and FF-CS-c, the flip in outflow spectrum shows up at smaller pump values in FF-NR-c than in FF-CS-c. The outflow spectrum is gradually decreasing in FF-NR-c with a large 8.2-fold rate to 6.7 nm linewidth, which is three-times smaller than in NF-NR-c (Fig. 2e). In contrast in case of FF-CS-c real narrowing occurs only at the largest pump intensity, however due to the largest 10.9-fold decrease from an intermediate passive linewidth the reached 3.9 nm FWHM is five-times smaller than in NF-CS-c, and is the smallest among the inspected systems (Table S1). The FF systems show larger bandwidth compared to their NF counterparts at small pumps, but this relation is reserved at the largest inspected pump intensity, proving the advantage of these systems in the far-field.The probe signal is not only amplified but it is redistributed in polar angle as well (Fig. 2f). The amplification is larger in FF-c than in NF-c nanoresonators, however the degree of redistribution is similar. The difference between FF-NR-c and FF-CS-c intensity is relatively smaller compared to the NF optimization.

By comparing the NF-c and FF-c nanoresonators one can conclude that the population inversion saturation is faster (slower) in the NR and CS nanoresonators optimized to maximize near-field enhancement (far-field out-coupling) (Fig S1, S2/a).

At the pump wavelength the $\varepsilon_{real}(\omega_a)$ uniformly does not modify, $\varepsilon_{imag}(\omega_a)$, as well as the $n(\omega_a)$ and $\kappa(\omega_a)$ slightly more rapidly (slowly) decreases in the nanoresonators optimized to maximize the near-field enhancement (far-field out-coupling) (Fig. S1, S2/b, d).

The $\varepsilon_{real}(\omega_e)$ and $n(\omega_e)$ more rapidly (slowly) decreases, whereas the $\varepsilon_{imag}(\omega_e)$ and $\kappa(\omega_e)$ modifies its decrease from faster to slower (from slower to faster), in the nanoresonators optimized to maximize the near-field enhancement (far-field out-coupling) (Fig. S1, S2/c, e). The $\varepsilon_{imag}(\omega_e)$ and $\kappa(\omega_e)$ are smaller (larger) in NF-CS-c (FF-CS-c) throughout the whole inspected interval and crosses zero earlier at $E_{sat}^{local}$: 0.56 (later at $E_{sat}^{local}$: 0.73), whereas in NF-NR-c (FF-NR-c) zero crossing occurs at $E_{sat}^{local}$: 0.52 (at $E_{sat}^{local}$: 0.62) but their relation modifies at $E_{sat}^{local} \sim 4$ (Table S1).

In NF-c (FF-c) nanoresonators the local field increases more rapidly (slowly) to larger (smaller) value before saturation (Fig. 1-to-2/a). The absorptance in gold is very similar and it is smaller (larger) in NF-c (FF-c) nanoresonators. The absorptance in the gain is smaller (larger) throughout the complete pump interval in NF-c (FF-c) nanoresonators (Fig. 1-to-2/b). For the NF-c systems the outflow remains smaller than the total power throughout the complete inspected pump interval, whereas in the FF-c systems the outflow overrides it above a certain outflow-threshold of the pump (Fig. 1-to-2/c).

The absorption cross-section of smaller (larger) maximal absolute value remains positive (becomes negative at a certain ACS-threshold) in NF-c (FF-c) nanoresonators. It is monotonous for NF-CS-c and FF-NR-c, whereas it exhibits an extremum for NF-NR-c and FF-CS-c. The scattering and the extinction cross-section exhibit a significantly slower (faster) monotonous increase to smaller (larger) value for the NF-NR (FF-NR). Both the scattering and extinction cross-section monotonously increases for NF-CS-c, whereas the scattering cross-section monotonously increases but the extinction cross-section exhibits a maximum for FF-CS-c (Fig. 1-to-2/d).

The average **E**-field increase is faster (slower) and the FWHM less (more) rapidly decreases in NF (FF) nanoresonators (Fig. 1-to-2/d). The outflow gradually decreases (flips above a certain threshold) by increasing the pump, and the FWHM significantly less (more) rapidly decreases (Fig. 1-to-2/e). The far-field lobes are smaller (larger) and are more (less) dependent on the nanoresonator type.

*Nanorod and core-shell mediated spasing: NF-c\* systems*

By increasing the concentration in the [$3*10^{25}$ m$^{-3}$, $3*10^{26}$ m$^{-3}$] interval the population inversion saturation behavior of NF-NR-c\* is intermediate in between NF and FF-NR-c, whereas it most rapidly increases and saturates in NF-CS-c\* (Fig. S1-3/a). This already indicates that the NF-NR-c\* is a kind of intermediate system, whereas the NF-CS-c\* deviates from both of NF and FF-CS systems.

In NF-NR-c\* in case of $5*10^{25}$ m$^{-3}$ concentration the saturation occurs at $E_{sat}^{local}$: 3.67 ($E_{sat}^{pump}$: 0.08), comparison of the internal local and external pump **E**-field strengths reveals 45.88-fold near-field enhancement at the pump wavelength, which is intermediate with respect to the enhancements achieved in NF-NR-c and FF-NR-c (Fig. 2a, Table S3). The NF-NR-c\* approximates the saturation dynamics of *FF-c at $E_{sat}^{local}$=6.* In NF-CS-c\* in case of $8*10^{25}$ m$^{-3}$ concentration the saturation occurs at $E_{sat}^{local}$: 2.79 ($E_{sat}^{local}$: 0.11), comparison of the internal local and external pump **E**-field strengths reveals 25.36-fold near-field enhancement at the pump wavelength, which is smaller than the enhancement reached in either of NF-CS-c or FF-CS-c (Fig. 2a, Table S1). The saturation arises at a significantly smaller $E_{sat}^{local}$ in NF-CS-c\* than in NF-NR-c\* moreover the saturation process is faster. This relation is similar (reversed) with respect to the NF (FF) counterpart nanoresonators.

In both NF-NR-c\* and NF-CS-c\* at the pump wavelength the difference with respect to the passive system becomes significantly larger in $\varepsilon_{im}(\omega_a)$, as well as in $n(\omega_a)$ and $\kappa(\omega_a)$ (Fig. S1-3/b, d, Table S1).

Their values become significantly larger and their decrease is considerably faster than in either of NF-c or FF-c counterparts. This causes a larger detuning and larger absorption as well.

In both of the NF-NR-c* and NF-CS-c* at the probe wavelength the significantly larger difference in the real parts ($\varepsilon_{real}(\omega_e)$ and $n(\omega_e)$) results in larger detuning, however the real parts decrease faster, than in either counterpart NF-c or FF-c nanoresonators. The more rapidly decreasing imaginary parts ($\varepsilon_{imag}(\omega_e)$ and $\kappa(\omega_e)$) result in a pronounced negative absorption, and the imaginary parts decrease through a wider interval, than in either of the counterpart NF-c or FF-c nanoresonators (SFig. 3c, e, Table S1). The values of $\varepsilon_{imag}(\omega_a)$, $n(\omega_a)$, $\kappa(\omega_a)$, $\varepsilon_{real}(\omega_e)$ and $n(\omega_e)$ are larger, whereas $\varepsilon_{imag}(\omega_e)$ and $\kappa(\omega_e)$ are smaller in NF-CS-c* than in NF-NR-c* (Fig. S1-3/ b, e). All optical parameter intervals are wider in case of NF-CS-c* than in NF-NR-c* (Table S1).

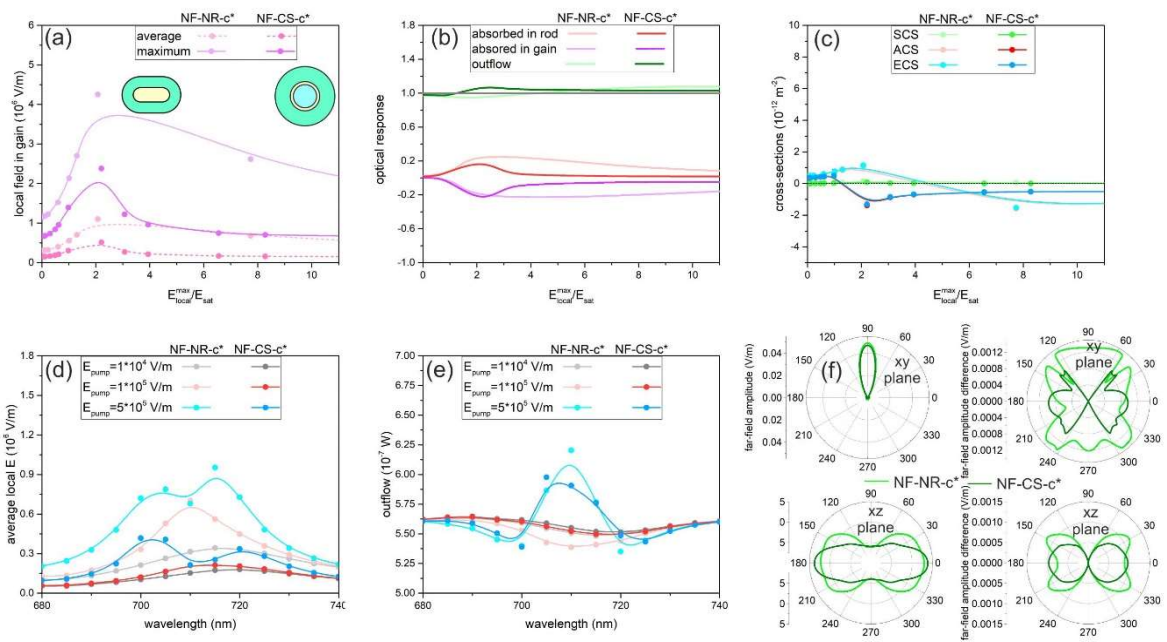

Figure 3. Pump dependent optical properties and response of NF-c optimized NF-NR-c* and NF-CS-c* nanoresonators with increased dye concentration: (a) average and maximum of the enhanced local probe electric field in the gain medium, (b) absorptions in the metal nanorod and shell, and in the gain medium, and the far-field emission, (c) optical cross-sections as a function of pump amplitude. Spectra of the (d) average near-field and (e) far-field power outflow at different pump amplitudes (f) polar angle distribution in the far-field. Inset: geometry of nanoresonators optimized to achieve the largest near-field enhancement with increased dye concentration (NF-NR-c*).

In NF-NR-c* the [$E_{sat}^{local}$: 0.65-2.07] region of linear near-field enhancement is contracted, a larger lasing threshold values ($E_{sat}^{local}$: 0.45 ($E_{sat}^{pump}$: 0.013)) are reached with a larger (4.84 10$^5$ (1.88 10$^6$ V/m)) slope efficiency in the average (maximal) probe **E**-field. At small and large pumps an intermediate value of the average probe **E**-field is achieved, however, instead of a saturation, a rapid monotonous near-field decrease occurs after a global maximum at $E_{sat}^{local}$ =2.07 (Fig. 3a, Table S1).

In NF-CS-c* the [$E_{sat}^{local}$: 0.30-2.20] region of linear near-field enhancement contracts, but smaller lasing threshold values ($E_{sat}^{local}$: 0.29 ($E_{sat}^{pump}$: 0.011)) are reached with a larger (1.87 10$^5$ (8.83 10$^5$ V/m)) slope efficiency in the average (maximal) probe **E**-field (Fig.3a, STable 1). At small and large pumps the smallest average **E**-field is achieved, but similarly to the NF-NR-c*, instead of a saturation, a rapid near-field decrease occurs after a global maximum at $E_{sat}^{loc}$ =2.2. In comparison, the lasing threshold is larger, and the slope efficiency as well as the achieved maximal and average local probe **E**-field is also larger in NF-NR-c* than in NF-CS-c*. Another significant difference is the faster decrease of the near-field in NF-CS-c*, when the pump is further increased.

The lasing threshold is the largest (smallest) in NF-NR(CS)-c*, however, the slope efficiency is the highest and in near-field enhancement both systems override the counterpart NF-NR-c and NF-CS-c nanoresonators at their maximum. The onset of gain occurs at a gain-threshold of $E_{sat}^{local}$: 0.51 ($E_{sat}^{pump}$: 0.011) in NF-NR-c* and $E_{sat}^{local}$: 0.51 ($E_{sat}^{pump}$: 0.02) in NF-CS-c*, respectively. The onset of gain occurs at smaller $E_{sat}^{pump}$ in case of NF-NR-c* but at the same $E_{sat}^{local}$ value in both nanoresonators. The initial and final deviation of NF-CS-c* is larger in all values and decrease faster, than in NF-NR-c* (Fig. 3b).

Correspondingly, in both of NF-NR-c* and NF-CS-c* the gold absorptance has similar (smaller) initial value compared to counterpart NF-c (FF-c) systems, but more rapidly increases and exhibits a maximum value that overrides the absorptance in FF-c systems as well, then starts to decrease. The gain exhibits a similar tendency as the gold absorptance, but with a slightly larger absolute value, which promotes the far-field power out-coupling. The decrease is more rapid in NF-CS-c* than in NF-NR-c*, however, the reached maximal gain is larger in NF-CS-c* nanoresonator (Fig. 3b).

On the far-field response an enhanced outflow appears, which reveals that the energy of the probe can be enhanced and out-coupled as well (Fig. 3b). This is due to that the smaller gain is accompanied by an even smaller absorptance in the metal rod (shell), as a result the loss is moderated as well. However, in NF-NR-c* this out-coupling occurs at larger pump than in FF-NR-c, and the out-coupled power remains smaller. In contrast, in NF-CS-c* the out-coupling arises at a smaller pump than in FF-CS-c, and its amplitude reaches larger values. In NF-NR-c* and NF-CS-c* the initial values of outflow are similar, but in case of NF-CS-c* it begins to increase earlier and faster. The out-coupling is achieved at considerably smaller pump ($E_{sat}^{loc}$=1.24) in NF-CS-c* than in NF-NR-c* ($E_{sat}^{loc}$=4.29), however the maximal out-coupled power reached at ($E_{sat}^{loc}$=2.2) is almost equal to that in NF-NR-c* at $E_{sat}^{loc}$=7.7 (Fig. 3b, Table S1).

Based on the pump dependent cross-section values both in NF-NR-c* and NF-CS-c* the scattering cross-section is relatively small and exhibits a maximum at $E_{sat}^{loc}$ =2.1 ($E_{sat}^{loc}$ =2.2). In NF-NR-c* (NF-CS-c*) the absorption cross-section also shows a maximum at $E_{sat}^{loc}$ = 2.1 ($E_{sat}^{loc}$ =1.0), then it becomes negative at $E_{sat}^{loc}$ =4.24 (1.26) as the pump intensity increases (Fig. 3c). The extinction cross-section is governed by the absorption cross-section, hence in contrast to the NF-NR-c and NF-CS-c it also exhibits a maximum than becomes negative as at $E_{sat}^{loc}$ =4.53 (1.29). Both the absorption and extinction reach the minimum at pump values corresponding to the maximal outflow. Comparison of NF-CS-c* to NF-NR-c* shows that the scattering cross-section is similarly small, the initial absorption cross-section is also smaller moreover it becomes negative more rapidly and exhibits a narrower dip of a similar negative value in NF-CS-c*, whereas the extinction cross-section is analogously determined by the absorption cross-section.

The near-field spectra in NF-NR-c* (NF-CS-c*) indicate that the large refractive index shift at smaller pump values makes the optimized system non-resonant around the emission wavelength, i.e. the peak redshifts to ~715 nm (~720 nm). Compared to counterpart NF and FF nanoresonators, the near-field increases most rapidly from an intermediate level, exhibits a narrowing but splits at large pump intensities. In case of very intense pumping an additional peak emerges again around 705 nm (700 nm) (with neighboring dips) on the near-field spectra (Fig. 3d). The only difference is that the splitting results in a global maximum at larger (smaller) wavelength in NF-NR-c* (NF-CS-c*) and the intensity of the spectra are always smaller in NF-CS-c*. In NF-NR-c* and NF-CS-c* the linewidths of the near-field spectra are similar, however caused by the smaller 3.2-fold linewidth narrowing the final linewidths of ~14.5 nm remain larger compared to the values of NF-c counterparts (Fig. 3d, Table S1). This indicates that the NF systems with post-optimization increased concentration are not optimal in the near-field. Although, the NF-CS-c* shows no linewidth narrowing at small pump intensities, at the largest pump its near-field spectral bandwidth is comparable to that of NF-NR-c*.

The far-field spectrum not only flips, it exhibits a narrowing as well, but the FWHM decrease is slightly (significantly) slower in NF-NR-c* (NF-CS-c*) (Fig. 3e). At the highest inspected pump the maximum on the far-field spectrum is coincident with 710 nm in NF-NR-c*, whereas it is still shifted to 705 nm in NF-CS-c* and shows an asymmetry (as a sign of coupled modes co-existence). The dips in NF-NR-c* spectra are deeper and are closer, whereas the peak is more intense. In its outflow spectrum NF-NR-c* shows significant 1.8-fold narrowing only at the largest inspected pump intensity (Fig. 3e, Table S1). According to the larger rate of FWHM decrease, the 15.5 nm linewidth is smaller than in NF-NR-c. Similarly, in NF-CS-c*, the 1.9-fold decrease is larger, as a result the reached minimal 16.3 nm linewidth is smaller than the values in NF-CS-c counterpart. Despite the larger rate of narrowing of NF-CS-c* the linewidth can be still slightly smaller in NF-NR-c* according to the narrower passive nanoresonator bandwidth. At large intensity the increased concentration becomes advantageous, but spectral bandwidths in NF-c* are intermediate with respect to the counterpart NF and FF active systems. This is in accordance with the pump intensity dependence of the outflow, which can be enhanced with increased concentration (Fig. 3b).

According to the polar diagram of far-field emission, NF-NR-c* is a unique system in the sense that it is capable of amplifying the probe beam along the direction of propagation, hence the redistribution in polar angle becomes less inhomogeneous compared to the NF-c and FF-c counterparts. There is no significant difference between the radiation patterns of NF-CS-c* and NF-CS-c despite the enhanced power outflow. Similarly to NF and FF optimized nanoresonators the outcoupling into the far-field is more efficient in NF-NR-c* than in NF-CS-c*.

These unique transitions indicate, that by changing considerably one of the configuration parameters, that is the dye concentration in present case, the optical response may undergo a significant modification as well.

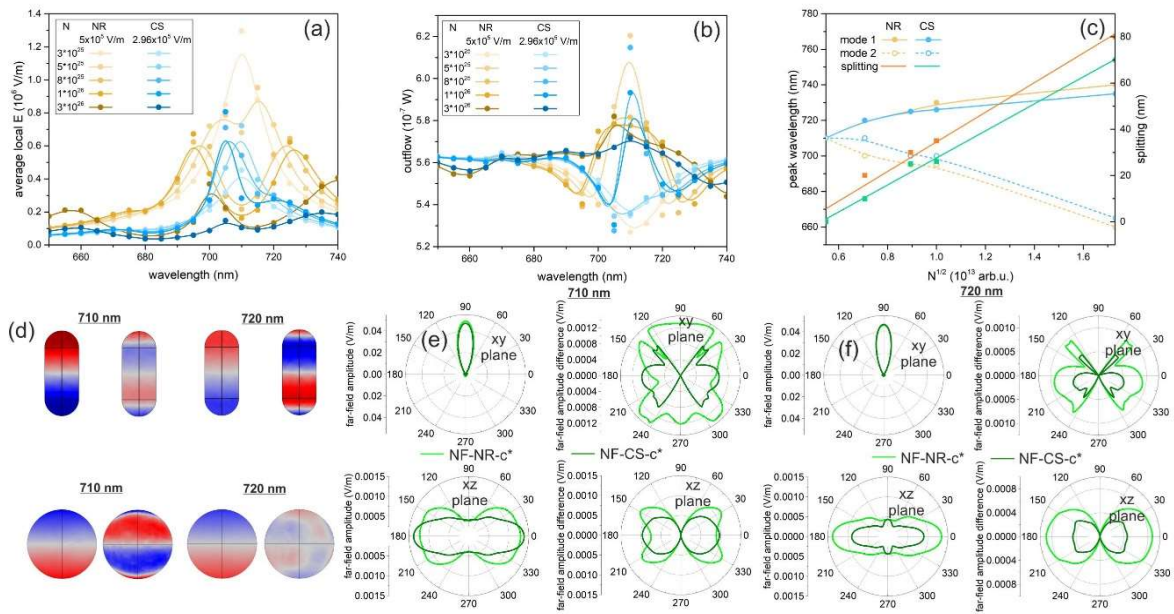

Figure 4. Effect of the concentration modification: (a) near-field spectra, (b) far-field spectra, (c) peak wavelengths and their splitting, (d) charge distribution and (e, f) far-field polar diagram at the (d, e) central peak (710 nm) and (f) one of the dips, in case of 5*10$^{25}$ m$^{-3}$ and 8*10$^{25}$ m$^{-3}$ concentration of dye at three different wavelengths for NF-NR-c* and NF-CS-c*.

The near- and far-field spectra at the pump intensities, which result in the maximal outcoupling in NF-NR-c* and NF-CS-c*, indicate that there is a strong interaction between different co-existing modes (Fig. 4a, b). Undoubtedly, the spectra collected by increasing the concentration from 3*10$^{25}$ m$^{-3}$ to 3*10$^{26}$ m$^{-3}$ reveals that at intermediate concentration (5*10$^{25}$ m$^{-3}$ and 8*10$^{25}$ m$^{-3}$) on the far-field spectral response of NF-NR and NF-CS a peak appears in between two dips (Fig. 4a, b).

The spectral distance of these dips is proportional to the $\sqrt{N}$ at intermediate concentrations, which reveals that two modes are strongly coupled in the active nanoresonator (Fig. 4c). The time-evolution of the charge separation at the peak and neighboring dips proves that although the dipolar mode is dominant at all extrema, at the peak a quadrupolar charge separation appears in an enhanced fraction of one time-cycle (Fig. 4d, Supplementary movies). However, this transient phenomenon becomes less dominant by increasing the concentration further. The symmetry of the polar angle distribution of the far-field emission pattern also indicates the co-existence of different modes (Fig. 4e, f).

**Conclusion**
The comparison of the nanorod and core-shell nanoresonator based NF-c and FF-c systems indicate, that the nanorod exhibits smaller lasing threshold, larger slope efficiency, and larger achieved **E**-field intensities due to larger near-field enhancement. The threshold in gain is slightly larger (smaller) in NF-NR (FF-NR), the optical cross-sections are larger in NF and FF-NR, and the FF-NR exhibit smaller threshold in outflow and ACS as well, than the counterpart CS nanoresonators. Accordingly the achieved peaks on the near-field spectra are larger, the outflow spectrum of NF-NR-c exhibits the sign of coupled modes, whereas the far-field spectrum flips at smaller pump and exhibits larger peaks in FF-NR-c. According to these results, the nanorod based resonators could be proposed for nanolasing in case of moderate concentration. However, the spectral bandwidth is smaller in the near-field in NF-CS-c and in the far-field in FF-CS-c with respect to the NR counterparts.

In case of the increased concentration the slope efficiency and achieved **E**-field intensity remain larger for the nanorod, but the lasing threshold becomes smaller for the core-shell nanoresonator. In addition to this the gain-threshold is equal, moreover the outflow, ACS and ECS-threshold also become smaller for the NF-CS-c*. The near-field and far-field is typically slightly smaller, but there is a pump interval ($E_{sat}^{loc}: 1.5 - 6$), where the outflow from the NF-CS-c* is significantly larger than from NF-NR-c*. In addition to this the IQE and EQE is also considerably larger in NF-CS-c*. Based on the comparison of the spectral, near-field and far-field responses at the pump intensities corresponding to outflow maxima, at larger concentration the core-shell nanoresonators become more competitive. The advantage of the NF-NR-c* is that it promotes the redistribution of the emission. In line-width narrowing both nanoresonators are weaker in the near-field, and intermediate in the fa-field, the NF-NR-c* (NF-CS-c*) exhibits narrower bandwidth in the near- (far-field), respectively.

In both types of nanoresonators the increase of concentration allows the transition to spasing, which requires a configuration, where the ECS crosses zero, and the gain is converted into far-field outflow. To the best of our knowledge this is the first case, when the strong-coupling of time-competing dipolar and quadrupolar modes is uncovered, and their role in facilitating the lasing-to-spasing transition is proven.

**Acknowledgements**
This work was supported by the National Research, Development and Innovation Office (NKFIH) project "Optimized nanoplasmonics" (K116362); "Ultrafast physical processes in atoms, molecules, nanostructures and biological systems" (EFOP-3.6.2-16-2017-00005); the University of Szeged Open Access Fund (4810).

**References**

[1] Huang, C.; Neretina, S.; El-Sayed, M. A. Gold Nanorods: From Synthesis and Properties to Biological and Biomedical Applications. *Adv. Mater.* **2009**, *21(48),* 4880-4910.

[2] Prodan, E.; Radloff, c.; Halas, N. J.; Nordlander, P. A hybridization model for the plasmon response of complex nanostructures. *Science* **2003**, *302(5644)*, 419–422.

[3] Jain, P. K.; Lee, K. S.; El-Sayed, I. H.; El-Sayed, M. A. Calculated Absorption and Scattering Properties of Gold Nanoparticles of Different Size, Shape, and Composition: Applications in Biological Imaging and Biomedicine. *J. Phys. Chem. B* **2006**, *110(14)*, 7238-7248.

[4] Averitt, R. D.; Westcott, S. L.; Halas, N. J. Linear optical properties of gold nanoshells. J*. Opt. Soc. Am. B* **1999**, *16(10)*, 1824-1832.

[5] Schuller, J. A.; Barnard, E. S.; Cai, W.; Jun, Y. C.; White, J. S.; Brongersma, M. L. Plasmonics for extreme light concentration and manipulation. *Nat. Mater.* **2010**, *9*, 193-204.

[6] Szenes, A.; Bánhelyi, B.; Szabó, L. Zs.; Szabó, G.; Csendes, T.; Mária, Cs. Enhancing Diamond Color Center Fluorescence via Optimized Plasmonic Nanorod Configuration. *Plasmonics* **2017**, *12*, 1263-1280.

[7] Szenes, A.; Bánhelyi B.; Szabó, L. Zs.; Szabó, G.; Csendes, T.; Csete, M. Improved emission of SiV diamond color centers embedded into concave plasmonic core-shell nanoresonators. *Sci. Rep.* **2017**, *7*, 13845.

[8] Oulton R. F.; Sorger, V. J.; Zentgraf, T; Ma, R-M.; Gladden, C.; Dai, L.; Bartal, G.; Zhang, X. Plasmon lasers at deep subwavelength scale. *Nature* **2009**, *461*, 629-632

[9] Ma, R-M.; Oulton, R. F.; Sorger, V. J.; Zhang, X. Plasmon lasers: coherent light source at molecular scales. *Laser Photonics Rev.* **2012**, *7(1)*, 1-21.

[10] Pang, L.; Freeman, L. M.; Chen, H. M.; Gu, Q.; Fainman, Y.; Plasmonics in Imaging, Biodetection, and Nanolasers. In *Handbook of Surface Science*, 1st ed.; Richardson, N.V., Holloway, S.; Elsevier: Amsterdam, NED, 2014; Volume 4, pp. 399-428.

[11] Thang, T; Shan, F. Development and Application of Surface Plasmon Polaritons on Optical Amplification. *J. Nanomater.* **2014**, *17*, 1-16.

[12] Chua, S-L.; Zhen, B.; Lee, J.; Bravo-Abad, J.; Shapira, O.; Soljačić, M. Modeling of threshold and dynamics behavior of organic nanostructured lasers. *J. Mater. Chemistry C* **2014**, *2*, 1463-1473.

[13] Marani, R.; D'Orazio, A.; Petruzzelli, V.; Rodrigo, S. G.; Martín-Moreno, L.; García-Vidal, F. J. Gain-assisted extraordinary optical transmission through periodic arrays of subwavelength apertures. *New J. Phys.* **2012**, *14*, 013020.

[14] Fietz, C.; Soukoulis C. M. Finite element simulation of microphotonic lasing system. *Opt. Express* **2012**, *20(10)*, 11548-11560.

[15] Pickering, T.; Hamm, J. M.; Page, A.F.; Wuestner, S.; Hess, O. Cavity-free plasmonic nanolasing enabled by dispersionless stopped light. *Nat. Commun.* **2014**, *5*, 4972.

[16] Fang, M.; Huang, Z.; Koschny, T.; Soukoulis C. M. Electrodynamic Modeling of Quantum Dot Luminescence in Plasmonic Metamaterials. *ACS Photonics* **2016**, *3(4)*, 558-563.

[17] Wuestner, S.; Pusch, A.; Tsakmakidis, K. L.; Hamm, J. M.; Hess, O. Gain and plasmon dynamics in active negative-index metamaterials. *Philos. T. R. Soc. A* **2011**, *369(1950)*, 3525-3550.

[18] Wuestner, S.; Hamm, J. M.; Pusch, A.; Renn, F.; Tsakmakidis, K. L.; Hess, O. Control and dynamic competition of bright and dark lasing states in active nanoplasmonic metamaterials. *Phys. Rev. B* **2012**, *85*, 201406.

[19] Hess, O.; Pendry, J. B.; Maier, A.; Oulton, R. F.; Hamm, J. M.; Tsakmakidis, K. L. Active nanoplasmonic metamaterials. *Nat. Mater.* **2012**, *11*, 573-584.

[20] Kim, M.; Oh, S. S.; Hess, O.; Rho, J. Frequency-domain modelling of gain in pump-probe experiment by an inhomogeneous medium. *J. Phys. Condens. Matter* **2018**, *30*, 064003.

[21] Hernández-Pinilla, D.; Cuerda, J.; Molina, P.; Ramírez, M. O.; Bausá, L. E. Spectral Narrowing in a Subwavelength Solid-State Laser. *ACS Photonics* **2019**, *6(9)*, 2327-2334.

[22] Noginov, M. A.; Thu, G.; Belgrave, A. M.; Bakker, R.; Shalaev, V.M.; Narimanov, E. E.; Stout, S.; Herz, E.; Suteewong, T.; Wiesner, U. Demonstration of a spaser-based nanolaser. *Nature*, **2009**, *460*, 1110-1112.

[23] Li, X. F.; Yu, S. F. Design of low-threshold compact Au-nanoparticle lasers. *Opt. Lett.* **2010**, *35(15)*, 2535-2537.

[24] Meng, X.; Kildishev, A. V.; Fujita, K.; Tanaka, K.; Shalaev, V. M. Wavelength-Tunable Spasing in the Visible. *Nano Lett.* **2013**, *13(9)*, 4106-4112.

[25] Cuerda, J.; García-Vidal, F. J.; Bravo-Abad, J. Spatio-temporal Modeling of Lasing Action in Core–Shell Metallic Nanoparticles. *ACS Photonics* **2016**, *3(10)*, 1952-1960.

[26] Arnold, N.; Piglmayer, K.; Kildishev, A. V.; Klar, T. A. Spasers with retardation and gain saturation: electrodynamic description of fields and optical cross-sections. *Opt. Mater. Express* **2015**, *5(11)*, 2546-2577.

[27] Arnold, N.; Hrelescu, C.; Klar, T. A. Minimal spaser threshold within electrodynamic framework: Shape, size and modes. *Ann. Phys-Berlin* **2015**, *528*, 295-306.

[28] Tao, Y.; Guo, Z.; Sun, Y.; Shen, F.; Mao, X.; Wang, W.; Li, Y.; Liu, Y.; Wang, X.; Qu, S. liver spherical nanoshells coated gain-assisted ellipsoidal silica core for low-threshold surface plasmon amplification. *Opt. Commun.* **2015**, *335*, 580-585.

[29] Kristanz, G. V.; Arnold, N.; Kildishev, A. V.; Klar, T. A. Power Balance and Temperature in Optically Pumped Spasers and Nanolasers. *ACS Photonics* **2018**, *5(9)*, 3695-3703.

[30] Ning, C-Z. What is Laser Threshold?. *IEEE J. Sel. Top. Quant.* **2013**, *19(4)*, 1503604.

[31] Chow, W. W.; Jahnke, F.; Gies, C. Emission properties of nanolasers during the transition to lasing. *Light Sci. Appl.* **2014**, *3*, e201.

[32] Azzam, S. I.; Kildishev, A. V.; Ma, R-M.; Nincs, C-Z.; Oulton, R.; Shalaev, V. M.; Stockman, M.I.; Xu, J-L.; Zhang, X. Ten years of spasers and plasmonic nanolasers. *Light Sci. Appl.* **2020**, *9*, 90.

[33] Galanzha, E. I.; Weingold, R.; Nedosekin, D. A.; Sarimollaoglu, M.; Nolan, J.; Harrington, W.; Kuchyanov, A. S.; Parkhomenko, R. G.; Watanabe, F.; Nima, Z.; Biris, A. S.; Plekhanov, A. I.; Stockam, M. I.; Zharov, V. P. Spaser as a biological probe. *Nat. Commun.* **2017**, *8*, 15528.

[34] Ma, R-M.; Oulton, R. F. Applications of nanolasers. *Nat. Nanotechnol.* **2019**, *14*, 12-22.

[35] Song, P.; Wang, J-H.; Zhang, M.; Yang, F.; Lu, H-J.; Kang, B.; Xu, J-J.; Chen, H-Y. Three-level spaser for next-generation luminescent nanoprobe. *Sci. Adv.* **2018**, *4(8)*, eaat0292.

[36] Gao, Z.; Wang, J-H.; Song, P.; Kang, B.; Xu, J-J.; Chen, H-Y. Spaser Nanoparticles for Ultranarrow Bandwidth STED Super-Resolution Imaging. *Adv. Mater.* **2020**, *32(9)*, 1907233.